%% file: main.tex
\documentclass[runningheads]{llncs}
\usepackage[T1]{fontenc}
\usepackage{graphicx,verbatim}
\usepackage{amsmath,amssymb}

\usepackage[colorlinks=true,linkcolor=blue]{hyperref}

\begin{document}
\title{\textit{ImmunoDiff}: A Diffusion Model for Immunotherapy Response Prediction in Lung Cancer}


\author{Moinak Bhattacharya\inst{1}
\and
Judy Huang\inst{1}
\and
Amna F. Sher\inst{1}
\and
Gagandeep Singh\inst{2}
Chao Chen\inst{1}
\and
Prateek Prasanna\inst{1}
}
\authorrunning{Bhattacharya et al.}
%
\institute{Stony Brook University, NY, US \and
Columbia University, NY, US\\}




\maketitle              
\begin{abstract}
Accurately predicting immunotherapy response in Non-Small Cell Lung Cancer (NSCLC) remains a critical unmet need. Existing radiomics and deep learning-based predictive models rely primarily on pre-treatment imaging to predict categorical response outcomes, limiting their ability to capture the complex morphological and textural transformations induced by immunotherapy. This study introduces \textit{ImmunoDiff}, an anatomy-aware diffusion model designed to synthesize post-treatment CT scans from baseline imaging while incorporating clinically relevant constraints. The proposed framework integrates anatomical priors, specifically lobar and vascular structures, to enhance fidelity in CT synthesis. Additionally, we introduce a novel cbi-Adapter, a conditioning module that ensures pairwise-consistent multimodal integration of imaging and clinical data embeddings, to refine the generative process. Additionally, a clinical variable conditioning mechanism is introduced, leveraging demographic data, blood-based biomarkers, and PD-L1 expression to refine the generative process. Evaluations on an in-house NSCLC cohort treated with immune checkpoint inhibitors  demonstrate a 21.24\% improvement in balanced accuracy for response prediction and a 0.03 increase in c-index for survival prediction. Code will be released soon.

\end{abstract}

\input{1_introduction}
\input{2_method}
\input{3_results}
\input{4_conclusion}
\bibliographystyle{splncs04}
\bibliography{reference}
%




\end{document}

%% file: 1_introduction.tex
\section{Introduction}
Non-small cell lung cancer (NSCLC) is the most prevalent form of lung cancer, accounting for approximately 85\% of cases worldwide \cite{siegel2024cancer}. Despite advances in early detection and treatment, NSCLC remains a leading cause of cancer-related mortality, with a five-year survival rate of approximately 28\% in the United States. Immunotherapy, particularly immune checkpoint inhibitors (ICIs) targeting Programmed cell death protein/ligand 1 (PD-1/PD-L1), has been shown to significantly improve patient outcomes \cite{rizvi2015mutational,borghaei2015nivolumab,gandhi2018pembrolizumab,hellmann2018nivolumab}. However, response to immunotherapy varies widely, necessitating the development of predictive models that can anticipate treatment outcomes based on pre-treatment imaging and clinical data.\\
\textbf{Clinical Motivation:} Measuring immunotherapy response in lung computed tomography (CT) involves using criteria such as RECIST or iRECIST. Existing radiomics  and deep learning-based response prediction methods~\cite{khorrami2020changes,she2022deep,liao2024personalized,rakaee2024deep} primarily encode pre-treatment imaging data to predict these hard labels (progression vs stable disease vs response). This presents notable limitations. 
To comprehensively characterize treatment response, it is essential to analyze the phenotype changes brought about by the ICIs - specifically the structural and textural transformation of the tumor and the lymph nodes from pre- to post-treatment. Therefore, rather than relying solely on pre-treatment imaging for response prediction, a more informative approach would involve studying the feature mapping between pre- and post-treatment scans in an image-to-image translation manner \cite{kim2024adaptive,hung2023med,webber2024diffusion,xing2024cross}.\\
\textbf{Technical Motivation:} Using generative modeling, image-to-image translation can be effectively achieved through learned high-dimensional mappings between source and target domains. Diffusion models, in particular, provide a powerful framework for learning high-fidelity mappings between imaging domains \cite{ho2020denoising,rombach2022high,zhang2023adding} and have demonstrated their effectiveness in medical image generation \cite{chambon2022roentgen,pinaya2022brain,zhang2024diffboost,bhattacharya2024radgazegen,bhattacharya2024gazediff}. Controllable diffusion models allow for incorporation of control factors to guide image generation with desired structures and shapes \cite{huang2023composer,li2023gligen,zhao2023uni,qin2023unicontrol}. This approach naturally extends to the medical domain, where clinical constraints can enhance the fidelity of generated images. Notably, integrating such constraints has improved 3D lung CT synthesis \cite{xu2024medsyn} and enabled CXR generation using lung segmentation masks and radiomics texture filters \cite{bhattacharya2024radgazegen}. 
In order to generate CT image of the lung region, lung vessels and lobes can be used as anatomical controls as these are the main structures visible in a CT scan. Previous studies have identified lobar features as relevant biomarkers for various lung diseases \cite{lei2022lobe,nilssen2024distribution}. Additionally, vascular features, such as blood vessel morphology, have been shown to be valuable indicators for predicting treatment response \cite{alilou2022tumor}. Furthermore, several studies have demonstrated that cancer treatments induce structural changes in the vasculature \cite{shrestha2024current,lin2023visualizing}.
Similarly, key clinical variables in immuno-oncology, such as demographics, blood biomarkers, and PD-L1 status, have been used as clinical baselines for response prediction but have not been explored as controls to a diffusion model.\\
\textbf{Contributions.} Based on these insights, we design an anatomy-guided diffusion model to generate post-treatment CT images and fine-tune this model for downstream immunotherapy response and survival prediction tasks. The main contributions of this work are as follows: (1) We propose an anatomy-guided diffusion model, incorporating lobar and vascular structures as control embeddings to improve CT image synthesis. (2) To generate post-treatment CT images from pre-treatment scans, we introduce a clinical variable guidance mechanism, where demographics, blood biomarkers, and PDL-1 status
are integrated as additional controls alongside pre-treatment image embeddings. (3) We extract features from the clinical variable-guided diffusion model for immunotherapy response and survival prediction. To the best of our knowledge, this is the first anatomy-guided diffusion model for post-treatment CT image generation that integrates pre-treatment imaging and clinical variables as controls, providing a novel framework for both image synthesis and predictive modeling in NSCLC.

%% file: 2_method.tex
\section{Method}
\begin{figure}[t]
    \centering
\includegraphics[width=1\linewidth]{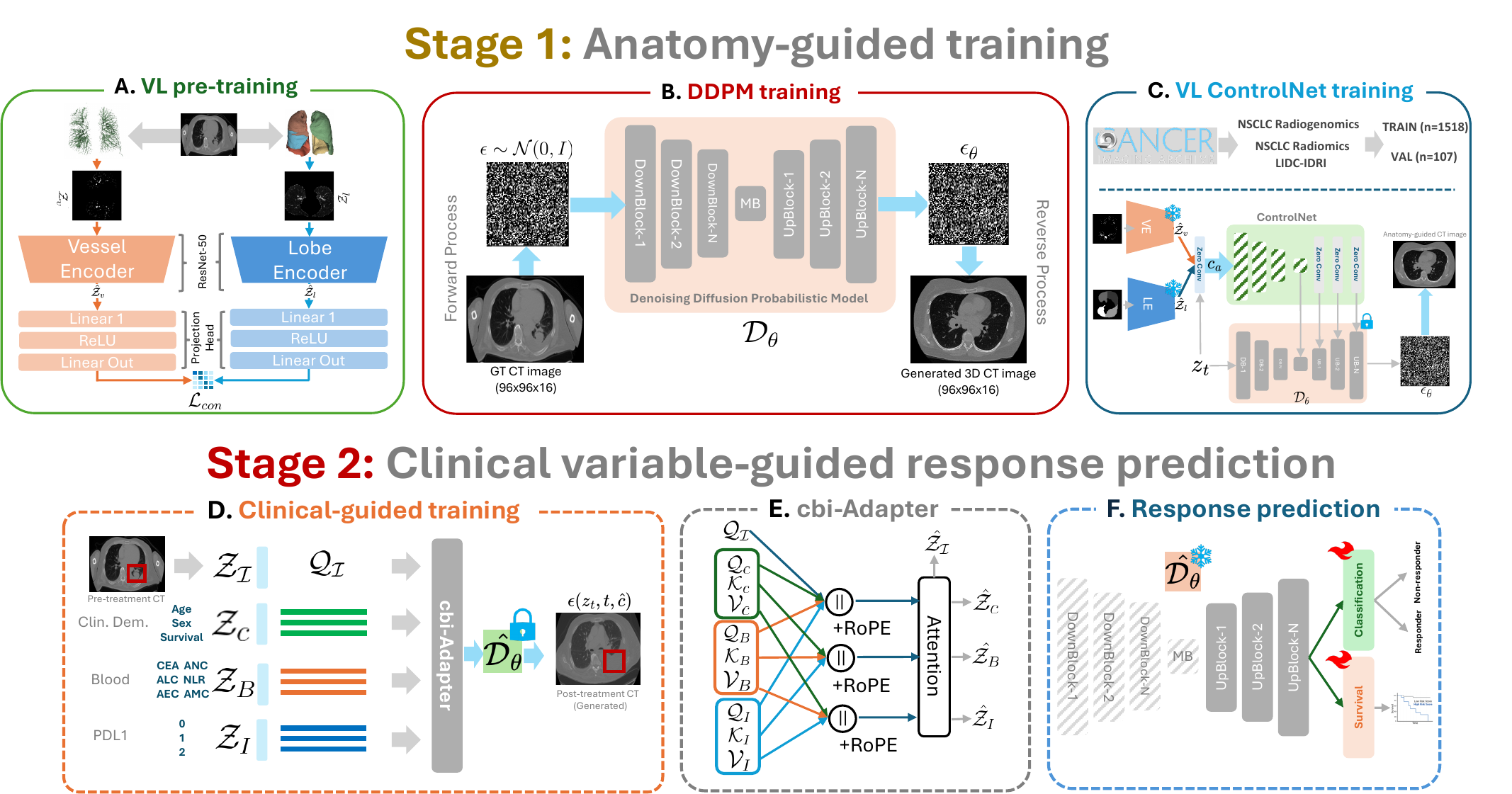}
\caption{\textbf{ImmunoDiff pipeline.}(A) Vessels and lobes masks
are pre-trained in a contrastive manner. (B) A DDPM model is trained on a large cohort of NSCLC patient CT images. (C) 
A ControlNet is trained with the vessel and lobe as controls. (D) Clinical variables are used as controls in addition with pre-treatment CT images to generate post-treatment CT images. (E) cbi-Adapter architecture.
(F) The trained diffusion model is used
for immunotherapy response prediction tasks (responder vs. non-responder classification and survival prediction).}
\label{fig:main}
\end{figure}

Figure \ref{fig:main} presents an overview of the proposed ImmunoDiff architecture. 
The architecture consists of two stages: (a) Anatomy-Guided Training (Stage 1) and (b) Clinical Variable-Guided Response Prediction (Stage 2). In Stage 1, we train a vessel and lobe (VL) encoder using a contrastive learning approach (Fig. \ref{fig:main}.A). Next, we train a denoising diffusion probabilistic model (DDPM) on a publicly available cohort of NSCLC patients from TCIA (details in Sec. \ref{results})(Fig. \ref{fig:main}.B). The trained encoders are then frozen and used as conditioning inputs for anatomy-guided ControlNet training (Fig. \ref{fig:main}.C). In Stage 2, we compute attention between various patient-specific clinical variables—including age, sex, blood parameters, and PD-L1 status—using the cbi-Adapter (Fig. \ref{fig:main}.E). This adapter is then used as a conditioning input for the ControlNet, which generates post-treatment CT images from pre-treatment scans and clinical variables (Fig. \ref{fig:main}.D). Additionally, features extracted from the 
upblocks of the U-Net are leveraged for immunotherapy response prediction (Fig. \ref{fig:main}.F). The following subsections provide a detailed discussion of each stage and its components.\\
\textbf{Preliminary.} Diffusion models first transfer the data distribution \textbf{x}$_0\sim q($\textbf{x}$_0)$ into Gaussian noise $\epsilon$ (\textit{forward process}) and learn to recover the data distribution by \textit{reversing} the forward process. The forward process is denoted by $q($\textbf{x}$_t|$\textbf{x}$_0)=\mathcal{N}($\textbf{x}$_t;\sqrt{\alpha_t}$\textbf{x}$_0,(1-\alpha_t)\textbf{I})$, where \textbf{x}$_t=\sqrt{\alpha_t}$\textbf{x}$_0+\sqrt{1-\alpha_t}\epsilon$, $\epsilon\sim\mathcal{N}(0,\textbf{I})$, $\alpha_t$ is the noise scale at each timestep $t\in T$, and $\mathcal{N}$ is the normal distribution. For the reverse process, the posterior distribution $q($\textbf{x}$_{t-1}|$\textbf{x}$_t,$\textbf{x}$_0)$ is learned where \textbf{x}$_{t-1}$ is recovered from \textbf{x}$_t$ by training a denoising U-Net \cite{ronneberger2015u} learning with parameter $\theta$. This model takes the noisy input $x_t$ and predicts the noise $\epsilon$ added in the forward process using the denoising mode $\epsilon_\theta($\textbf{x}$_t, t)$.
\subsection{Stage 1: Anatomy-guided training} 
In immuno-oncology, patient cohort data is often limited, and strict eligibility criteria further reduce the number of patients included in response prediction studies, resulting in small training datasets \cite{mahmoud2022machine,ventola2017cancer,wang2023cancer}. Training a diffusion model on limited-cohort datasets presents a significant challenge. This limitation can be mitigated by leveraging transfer learning through pre-training on large-scale publicly available datasets, ideally comprising the same disease domain. Here we propose a two-stage training strategy: first, pre-training the diffusion model using publicly available NSCLC datasets to learn a robust feature representations, followed by fine-tuning on a smaller cohort of patients treated with immunotherapy to adapt the model for treatment-specific transformations.

Another critical challenge in generative modeling of medical images is ensuring anatomical fidelity in the synthesized CT scans. To address this, we incorporate lung vasculature  and lobe structures as anatomical controls, guiding the generation process to maintain clinically relevant spatial and morphological integrity. Further, to enhance the structural coherence between vasculature and lobar segmentation, we employ a contrastive learning framework. This ensures that the synthesized images not only exhibit high visual fidelity but also preserve meaningful anatomical relationships critical for downstream clinical interpretation. The details are discussed next:\\
\textbf{VL pre-training.} In order to ensure the multi-anatomy synchronization, we first train the vasculature and lobe structures in a contrastive manner (Fig. \ref{fig:main}.A). The masked vascular and lobar regions are represented as $\mathcal{Z}_v$ and $\mathcal{Z}_l$, respectively. The vessel and lobe feature encoders are $\mathcal{F}_{v}$ and $\mathcal{F}_{l}$, and the
embeddings are $h_v=\mathcal{F}_v(\mathcal{Z}_v)$ and $h_l=\mathcal{F}_l(\mathcal{Z}_l)$, here $h\in\mathbb{R}^d$ is the output of the $AvgPool$ layer. The projection heads, $g_v(.)$ and $g_l(.)$ maps the encoded representation $h_v$ and $h_l$ to a space where the contrastive loss is calculated. This is shown as $\hat{\mathcal{Z}}_v=g(h_v)=W_2(ReLU(W_1.h_v))$ and same for $\hat{\mathcal{Z}}_l$. The contrastive loss $\mathcal{L}_a$ is defined as:
\begin{equation}
    \mathcal{L}_a = \frac{1}{N}\sum^N_{i=1}-\log\frac{\exp(sim(\hat{\mathcal{Z}}_{v_i},\hat{\mathcal{Z}}_{l_i})/\tau)}{\sum^N_{k=1}\exp(sim(\hat{\mathcal{Z}}_{v_i}, \hat{\mathcal{Z}}_{l_k})/\tau)}
\end{equation}
Here, $\tau$ is the temperature parameter which is set to $0.5$, $N$ is the total data size and $(v_i, l_i)$ are the positive vessel and lung pairs from the same patient and $(v_i, l_k)$ are the negative pairs from different patients.\\
\textbf{Training.} 
Now, the lobe and vessel masks are fed to the frozen $\mathcal{F}_l$ and $\mathcal{F}_v$ and the generated embeddings are used for conditioning the anatomy-guided diffusion model, similar to ControlNet \cite{zhang2023adding}. First a 
DDPM model, $\mathcal{D}_\theta$ is trained with $x_0$ which is sampled from the publicly available NSCLC dataset. 
The loss function $\mathcal{L}_\mathcal{D}(\theta)=\mathbb{E}_{t, x_0, \epsilon}\left[\left\Vert\epsilon-\epsilon_\theta(\sqrt{\alpha_t}x_0+\sqrt{1-\alpha_t}\epsilon, t)\right\Vert^2\right]$. 
This large pre-trained diffusion model is now locked and conditioned with lobe and vessel anatomies as controls. The vessel and lobe conditioning embeddings are represented as $\hat{\mathcal{Z}}_v$ and $\hat{\mathcal{Z}}_l$ respectively. These are fed as control $c_a$, where $c_a=\hat{\mathcal{Z}}_v\oplus\hat{\mathcal{Z}}_l$, to the locked $\mathcal{D}_\theta$ and trained with the loss function $\mathcal{L}_{CN}=\mathbb{E}_{z_1,t, \hat{c}_a, \epsilon\sim\mathcal{N}(0,1)}\left[\left\Vert\epsilon-\epsilon_\theta(z_t, t, \hat{c}_a)\right\Vert^2_2\right]$.
\subsection{Stage 2: Clinical variable-guided response prediction}
The vessel and lobe-guided training helps in generating anatomically accurate CT images. Now, in order to generate the post-treatment CT images, we need to condition the diffusion model with the pre-treatment CT images and the clinical variables. The clinical variables condition the generation of post-treatment images from the pre-treatment images based on 
PDL1-status, blood parameters and demographics of patients. Since, these variables are critical in predicting the treatment outcomes \cite{mbanu2022clinico}, we hypothesize that these can be used as clinically useful controls in post-treatment image generation.\\
\textbf{cbi-Adapter.} In Fig. \ref{fig:main}.E, we show the architecture of cbi-Adapter. The cbi-Adapter integrates embeddings from two conditions as inputs: clinical variables and pre-treatment CT images. 
The joint attention mechanism representing \textbf{c}linical demographics, \textbf{b}lood parameters, and \textbf{i}mmunotherapy biomarkers is collectively termed the \textbf{cbi}-Adapter. This integration enables unified controllable post-treatment CT image generation by improving the cross-clinical attention and the clinical-to-image domain translation. The embeddings from the pre-treatment CT image is represented as $\mathcal{Z}_\mathcal{I}$, and the clinical embeddings are represented as $\mathcal{Z}_C$, $\mathcal{Z}_B$, and $\mathcal{Z}_I$, respectively. The embeddings are mapped to query $\mathcal{Q}_i$, key $\mathcal{K}_i$ and value $\mathcal{V}_i$ features through linear transformations $L_i$, where, $i\in\{\mathcal{C}, B, I\}$. We apply 1-D Rotatory Position Embedding (RoPE) \cite{su2024roformer} to enhance the controllable precision of the clinical variables. RoPE computes relative clinical embedding similarity across the query, key and value vectors
, hence improving the spatial alignment of clinical variables and image embedding space. Given, \textbf{f}$\in\mathbb{R}^{|D|}$ is a feature vector of the $\{\mathcal{Q}_i, \mathcal{K}_i, \mathcal{V}_i\}$ at position $(h, w)$, the RoPE is computed as \textbf{f}$=[\mathcal{R}_h$\textbf{f}$_h||\mathcal{R}_w$\textbf{f}$_w]$, where $\mathcal{R}$ is the rotary matrix, h is height and w is width, defined as
\begin{equation}
    \mathcal{R}_{k\in\{h|w\}}=\left[\begin{smallmatrix}
        \cos(k\Theta_0) & -\sin(k\Theta_0) & \cdots & 0 & 0\\
        \sin(k\Theta_0) & \cos(k\Theta_0) & \cdots & 0 & 0\\
        \vdots & \vdots & \ddots & \vdots & \vdots\\
        0 & 0 & \cdots & \cos(k\Theta_l) & -\sin(k\Theta_l) \\
        0 & 0 & \cdots & \sin(k\Theta_l) & \cos(k\Theta_l)
        \end{smallmatrix}\right]; \Theta_i=b^{-i/|D|}
\end{equation}
Here, $b=1e5$ and $l=(|D|/4)-1$. Then, an attention mechanism attends to these parameters and maps to $\hat{\mathcal{Z}_i}$, represented as $\hat{\mathcal{Z}_i}=\prod_i Attn(\mathcal{Q}_i, \mathcal{K}_i, \mathcal{V}_i)$. This $\hat{\mathcal{Z}}_i$ is used as an additional conditioning to $\hat{\mathcal{D}}_\theta$ discussed next.\\
\textbf{Training.} Given the post-treatment CT image $z_1$, noise is progressively added to the image producing a noisy image $z_t$, where $t$ is the number of timesteps for which the noise is added (Fig. \ref{fig:main}.D). In addition to $z_0$, which is the pre-treatment CT image, the output embeddings from the cbi-Adapter, $\hat{\mathcal{Z}}_C$, $\hat{\mathcal{Z}}_B$ and $\hat{\mathcal{Z}}_I$ are fused as a unified condition to $\hat{\mathcal{D}}_\theta$. The embedding from the pre-treatment CT image, denoted as $\mathcal{Z}_\mathcal{I}$, is obtained by processing $z_0$ through a $SiLU(Conv2D(.))$ function, yielding the final embeddings $\hat{\mathcal{Z}}_\mathcal{I}$. The  The overall learning objective $\mathcal{L}_{final}$ is defines as
    $\mathcal{L}_{final}=\mathbb{E}_{z_1,t, \hat{c}, \epsilon\sim\mathcal{N}(0,1)}\left[\left\Vert\epsilon-\epsilon_\theta(z_t, t, \hat{c})\right\Vert^2_2\right]$, where $ \hat{c}=\hat{\mathcal{Z}_\mathcal{I}}+\lambda\sum^{(\mathcal{C}, B, I)}_i\hat{\mathcal{Z}_i}.$
Here, $\hat{c}$ is the weighted combination of the conditioning inputs to $\hat{D}_\theta$ and $\lambda$ is the weighting parameter which is set to $5e-2$.\\
\textbf{Treatment Response Assessment.} After training $\hat{\mathcal{D}_\theta}$, the features from the frozen U-Net are extracted and used for downstream clinical tasks such as immunotherapy response prediction and patient survival risk estimation (Figure \ref{fig:main}.E). \textit{a) Immunotherapy response prediction.} The features $\hat{\textbf{f}}$ from the upblocks of frozen $\hat{\mathcal{D}}_\theta$ are extracted and fed to a classification head for predicting binary labels for responders and non-responders. \textit{b) Survival Analysis.} Similar to the immunotherapy response prediction, the $\hat{\textbf{f}}$ is fed to a survival head where the hazard function is represented as $H(t)=\int_0^{T}h(u)du$.

%% file: 3_results.tex
\vspace{-0.08in}
\section{Results}
\label{results}

\input{tables/table_1}
\textbf{Dataset and Implementations.} 
For Stage 1, we use three publicly available datasets from The Cancer Imaging Archive (TCIA), namely NSCLC Radiomics \cite{aerts2014decoding}, NSCLC Radiogenomics \cite{bakr2018radiogenomic} and LIDC-IDRI \cite{armato2011lung}. The number of CTs in the training and validation set are 1518 and 107 respectively. The lung lobes were extracted using~\cite{hofmanninger2020automatic}, while the vessels were segmented using TotalSegmentator \cite{wasserthal2023totalsegmentator}. For immunotherapy response analysis, we use an in-house dataset where patients with Stage III-IV NSCLC are treated with ICI therapy (including pembrolizumab, nivolumab, durvalumab, and atezolizumab). For Stage 2, The total number of patients used are 74 and the training is done in a 5-fold cross-validation setting. The response was determined via RECIST 6-weeks post therapy.
There are a total of 19 responders (complete or partial response) and 55 non-responder patients (progression or stable disease). The clinical variables include:  demographics (Age, Sex, and Race), blood parameters (CEA (carcinoembryonic antigen), ANC (absolute neutrophil count), ALC (absolute lymphocyte count), NLR (neutrophil to lymphocyte ratio), AEC (absolute eosinophil count), and AMC (absolute monocyte count)) and PDL1 status. We crop the CT images to (96, 96, 16) shape with [3.0, 3.0, 0.5] spacing due to computational resource limitations. For training, we use Adam optimizer with a learning rate of $2.5e-5$ and batch size of 2. Training is done on a Quadro RTX 8000 (48 GB) GPU.

\begin{figure*}[t]
    \centering
        \includegraphics[width=.8\linewidth]{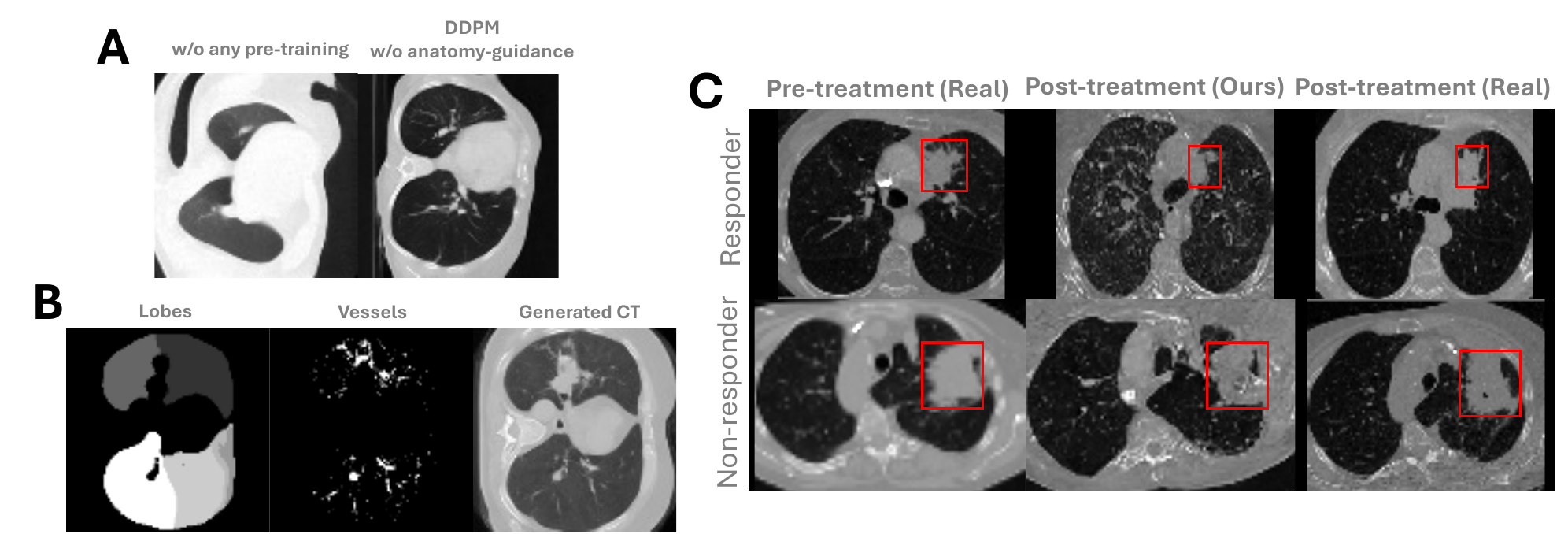}
    \caption{\textbf{Qualitative results.}  (A) CT images generated w/o pre-training and with pre-training but w/o anatomy-guided training. (B) 
    CT image generated using anatomy-guided training. Also shown are the lobes and vessels used as controls.
    (C) Post-treatment CT generated by ImmunoDiff for responder and non-responder cases. 
    (tumors shown in \textcolor{red}{red box}).
    }
    \label{fig:qualitative}
\end{figure*}

\input{tables/table_2}
\noindent\textbf{Immunotherapy Response Prediction.} In Table \ref{tab1}, we present the performance evaluation of ImmunoDiff for immunotherapy response classification in comparison to various baseline models. The baselines include conventional CNNs such as ResNet and DenseNet, as well as transformer architectures like Vision Transformer (ViT) and Swin Transformer. Additionally, we report results for these image encoders when integrated with clinical parameters ($+\mathcal{C}_X$). To ensure a comprehensive assessment, we also include several diffusion-based classification methods, including Zero-shot Diffusion Classifier (0-shot DC), DDPM for Anomaly Detection (DDPM-AD), and ControlNet, as baseline comparisons.
Our results demonstrate that ImmunoDiff outperforms all baseline models, achieving a 21.24\% improvement in Balanced Accuracy, a 7.32\% increase in F1-score, and a 13.8\% enhancement in Precision over the second-best baseline, while maintaining a comparable Recall of 0.75±0.12 to ViT (0.75±0.08), Swin Transformer (0.75±0.08), and DenseNet-121$+\mathcal{C}_X$ (0.75±0.08). These findings demonstrate the robustness of ImmunoDiff, particularly in handling class-imbalanced datasets.
In Table \ref{tab2}, we report the Concordance Index (c-index) for survival prediction. ImmunoDiff surpasses the second-best baseline by 0.03 in c-index.\\
\noindent\textbf{Ablations.} 
In Table \ref{tab3}, we present a detailed evaluation of the individual contributions of Stage 1 and Stage 2 across multiple performance metrics. Specifically, we assess image quality (IQ) using Maximum Mean Discrepancy (MMD) and Structural Similarity Index Measure (SSIM), immunotherapy response prediction using F1-score, Precision, and Recall, and survival prediction using the c-index.
Our results indicate that the combined approach (Stage 1+2, Ours) consistently outperforms the individual stages across all IQ and survival prediction metrics, as well as two out of three immunotherapy response prediction metrics. Notably, while Stage 1 alone achieves a slightly higher Precision score, it does not lead to superior overall performance. These ablation experiments confirm that neither anatomy-guided training nor cbi-Adapter alone is sufficient to generate clinically accurate and high-quality CT images. Instead, the integration of cbi-Adapter enhances both image quality and response prediction performance, highlighting the necessity of a synergistic approach in ImmunoDiff.\\
\noindent\textbf{Qualitative Analysis.} We present the qualitative results for anatomy-guided image generation and post-treatment CT generation in Fig. \ref{fig:qualitative}.
In Fig. \ref{fig:qualitative}.A, we compare the generated images without pretraining to those produced by a DDPM trained on a large public cohort without anatomy guidance (Stage 1.B). A radiologist ($8$ years of experience) reviewed these results, observing that the images generated by the DDPM trained on a large dataset exhibit improved fidelity compared to those generated without pretraining. This emphasizes the necessity of training with large-scale datasets to enhance image quality and realism.
Fig. \ref{fig:qualitative}.B, shows the CT images generated by Stage 1.C, which incorporates anatomical guidance using structures such as lobes and vessels. The generated images closely resemble the control images in terms of lobe and vessel structures, highlighting the effect of anatomy-guided training in preserving anatomical consistency.
Finally, in Fig. \ref{fig:qualitative}.C, we present the post-treatment CT generated by ImmunoDiff for different response groups. 
Our observations reveal that tumor size decreases in the generated post-treatment CT images for responder patients, whereas it remains stable or increases for non-responders, demonstrating ImmunoDiff’s ability to capture treatment response dynamics effectively.

%% file: tables/table_1.tex
\begin{table}[t]
\centering
\caption{Immunotherapy Response Prediction. Best results in \textbf{bold} and second best in \underline{underline}.}
\label{tab1}
\scalebox{0.8}{
\begin{tabular}{|c|c|c|c|c|}
\hline
- &  Bal. Acc. ($\uparrow$) & F1 ($\uparrow$) & Precision ($\uparrow$) & Recall ($\uparrow$)\\
\hline
ResNet-50\cite{he2016deep} & \underline{0.53±0.09} & \underline{0.69±0.13} & \underline{0.66±0.14} & 0.74±0.11\\
DenseNet-121\cite{huang2017densely} & 0.47±0.07 & 0.56±0.13 & 0.64±0.11 & 0.54±0.12\\
ViT-B/16\cite{dosovitskiy2020image} & 0.50±0.00 & 0.65±0.11 & 0.58±0.14 & \underline{0.75±0.08}\\
Swin-Base\cite{liu2021swin} & 0.50±0.00 & 0.65±0.11 & 0.58±0.14 & \underline{0.75±0.08}\\
\hline
ResNet-50+$\mathcal{C}_X$ & 0.49±0.10 & 0.22±0.22 & 0.26±0.27 & 0.30±0.18\\
DenseNet-121+$\mathcal{C}_X$ & 0.50±0.00 & 0.65±0.11 & 0.58±0.14 & \underline{0.75±0.08}\\
ViT-B/16+$\mathcal{C}_X$ & 0.50±0.00 & 0.29±0.25 & 0.23±0.22 & 0.41±0.25\\
Swin-Base+$\mathcal{C}_X$ & 0.50±0.00 & 0.10±0.04 & 0.06±0.03 & 0.24±0.08\\
\hline
0-shot DC\cite{li2023your} & 0.42±0.13 & 0.23±0.10 & 0.26±0.15 & 0.30±0.13\\
DDPM-AD\cite{wolleb2022diffusion} & 0.44±0.05 & 0.48±0.24 & 0.51±0.25 & 0.52±0.21\\
ControlNet\cite{zhang2023adding} & 0.43±0.12 & 0.40±0.09 & 0.61±0.10 & 0.38±0.05\\
\hline
Ours & \textbf{0.74±0.20} & \textbf{0.76±0.13} & \textbf{0.79±0.16} & \textbf{0.75±0.12}\\
\hline
\end{tabular}
}
\end{table}

%% file: tables/table_2.tex
\begin{table}[t]
\parbox{.33\linewidth}{

\centering
\caption{Survival Analysis.}
\scalebox{0.8}{
\label{tab2}
      \begin{tabular}{|c|c|c|}
      
        \hline
        - &  c-index($\uparrow$)\\
        \hline
        DDPM & 0.54±0.08\\
        CN & 0.52±0.14\\
        \hline
        Ours & \textbf{0.57±0.12}\\
        \hline
        \end{tabular}}

}
\hfill
\parbox{.7\linewidth}{

\centering
\caption{Ablation experiments.}
  \scalebox{0.72}{
      \label{tab3}
    \begin{tabular}{|c|c|c|c|c|c|c|}

    \hline
    - & \multicolumn{2}{c|}{IQ} & \multicolumn{3}{c|}{Response pred.} & Surv. \\
    \hline
    - &  MMD($\downarrow$) & SSIM($\uparrow$) & F1($\uparrow$) & Precision($\uparrow$) & Recall($\uparrow$) & c-index($\uparrow$)\\
    \hline
    Stage-1 & 1.0789 & 0.2256 & 0.70±0.03 & \textbf{0.82±0.07} & 0.72±0.05 & 0.47±0.04\\
    Stage-2 & 2.3528 & 0.2034 & 0.64±0.13 & 0.56±0.15 & 0.74±0.09 & 0.54±0.07\\
    \hline
    Ours & \textbf{0.0330} & \textbf{0.2979} & \textbf{0.76±0.13} & 0.79±0.16 & \textbf{0.75±0.12} & \textbf{0.57±0.12}\\
    \hline
    \end{tabular}}
}
\end{table}



%% file: 4_conclusion.tex
\vspace{-0.08in}
\section{Conclusion} In this work, we present ImmunoDiff, a novel clinical variable- and anatomy-guided diffusion model for immunotherapy response prediction. By incorporating vasculature and lobar information as anatomical controls, our approach enhances the generation of anatomically coherent CT images. Furthermore, conditioning the diffusion process on clinical variables— such as  immunotherapy-specific markers— enables the synthesis of diagnostically meaningful post-treatment scans from pre-treatment images. We also demonstrate the efficacy of the extracted features in predicting treatment response. \\
\textbf{Acknowledgments} This research was supported by the National Institutes of Health (NIH) grants R21CA258493-01A1 and R01CA297843. The content is solely the responsibility of the authors and does not necessarily represent the official views of the National Institutes of Health.